\documentclass[aps,prl,twocolumn,showpacs,superscriptaddress]{revtex4-1}  
\usepackage{graphicx}  
\usepackage{dcolumn}   
\usepackage{bm}        
\usepackage{braket}
\usepackage{exscale}                  
\usepackage[intlimits]{amsmath}       
\usepackage{bbm}
\usepackage{amsfonts}
\usepackage{amssymb,amscd}
\usepackage{color}
\usepackage{subfigure}
\usepackage[normalem]{ulem}
\usepackage[utf8]{inputenc}
\usepackage{hyperref}

\hypersetup{colorlinks=true,linkcolor=blue,citecolor=blue,urlcolor=black}

\usepackage{nicefrac}
\usepackage{xcolor}

\definecolor{lcolor}{rgb}{0.5,0,0}
\definecolor{citcolor}{rgb}{0,0.3,0.0}
\definecolor{azure}{rgb}{0.0, 0.5, 1.0}

\usepackage{comment}

\newcommand{\mbf}{\mathbf}

\newcommand{\pToFigs}{.}

\newcommand{\cl}{\text{cl}}

\newcommand{\nL}{\text{nL}}
\newcommand{\Lor}{\text{largeN}}

\newcommand{\mEff}{M}

\DeclareMathOperator{\sech}{sech}

\begin{document}

\title{Unraveling the nature of universal dynamics in $O(N)$ theories}

\author{Kirill Boguslavski}
\affiliation{Institute for Theoretical Physics, Technische Universit\"{a}t Wien, 1040 Vienna, Austria}
\affiliation{Department of Physics, University of Jyv\"{a}skyl\"{a}, P.O.~Box 35, 40014 Jyv\"{a}skyl\"{a}, Finland}
\author{Asier Pi\~neiro Orioli}
\affiliation{JILA, Department of Physics, University of Colorado, Boulder, Colorado 80309, USA}
\affiliation{Center for Theory of Quantum Matter, University of Colorado, Boulder, CO 80309, USA}


\begin{abstract}
Many-body quantum systems far from equilibrium can exhibit universal scaling dynamics which defy standard classification schemes.
Here, we disentangle the dominant excitations in the universal dynamics of highly-occupied $N$-component scalar systems using unequal-time correlators. While previous equal-time studies have conjectured the infrared properties to be universal for all $N$, we clearly identify for the first time two fundamentally different phenomena relevant at 
different $N$. We find all $N\geq3$ to be indeed dominated by the same Lorentzian ``large-$N$'' peak, whereas $N=1$ is characterized instead by a non-Lorentzian peak with different properties, and for $N=2$ we see a mixture of two contributions.
Our results represent a crucial step towards obtaining a classification scheme of universality classes far from equilibrium.
\end{abstract}

   
\maketitle



{\it Introduction.---}Universality constitutes a powerful tool to understand complex many-body systems.
A remarkable example are equilibrium phase transitions, where theories can be classified into universality classes based on only few system parameters 
\cite{goldenfeld2018lectures}.
Out of equilibrium, while a comprehensive picture is lacking, universal scaling phenomena have been found in turbulence~\cite{KolmogorovTurbulence}, coarsening~\cite{PhaseOrderingKinetics}, ageing~\cite{Calabrese_2005}, or driven-dissipative systems~\cite{DrivDissSys_SiebererPRL}.
In recent years, new far-from-equilibrium universality classes for isolated quantum systems have been theoretically identified~\cite{Berges:2008wm,Micha:2002ey,NowakPRB84,Berges:2012us,NowakPRA85,SextyGasenzerPLB2012,PineiroPRD92,WalzPRD97,ChantesanaPRA99,GasenzerMikheevPRA99,MoorePRD93,Berges:2016nru,Berges:2012iw,Berges:2013eia,Berges:2014bba,Boguslavski:2019fsb,Bhattacharyya:2019ffv,Dolgirev:2019udo,Mace:2019cqo}
which have recently started to be probed in cold-atom experiments~\cite{Prufer:2018hto,Erne:2018gmz,HadzibabicNature563,Prufer:2019kak,Zache:2019xkx}.
These universality classes can encompass vastly different theories such as gauge and scalar theories~\cite{Berges:2014bba,Berges:2019oun}, or relativistic and non-relativistic theories~\cite{PineiroPRD92}.
These unexpected connections raise the question of what the relevant physics behind the observed universality is.

\begin{figure}[t!]
\centering
\includegraphics[width=\columnwidth]{\pToFigs/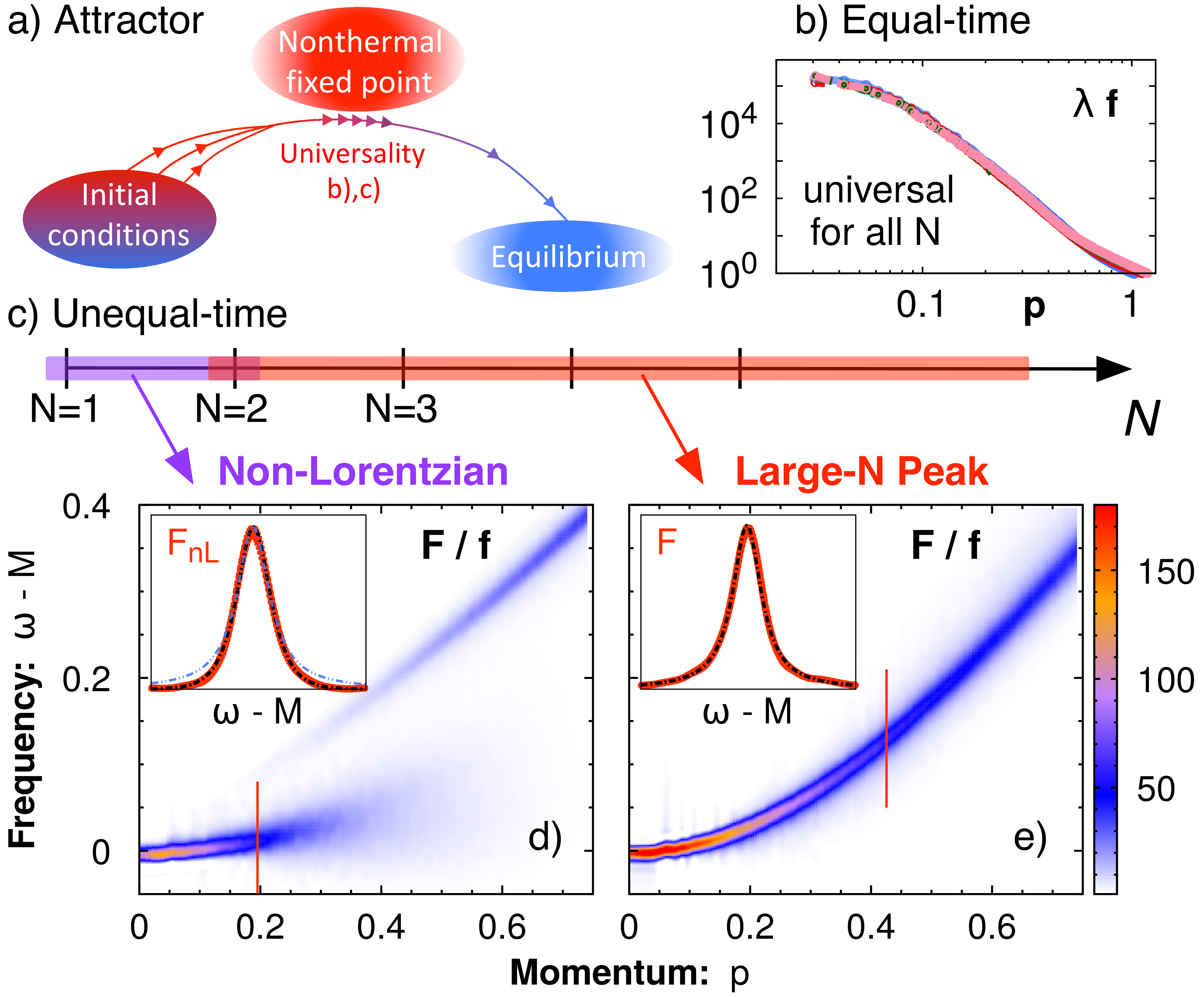}
\caption{\emph{Top:} a) Typical thermalization scenario for large initial occupancies, the system approaches an attractor during its evolution. 
b) The distribution function $f(t,p)$ at low momenta close at the attractor for $N = 1,2,3,4,8$ at times $t = 600,600,750,1000,2000$, respectively. 
c) Illustration of universality classes for different $N$. 
\emph{Bottom:} Statistical function $F(\tau,\omega,p)/f(\tau,p)$ for $N=1$ (left, d) and $N=8$ (right, e)  at $\tau = 1000$. Red lines correspond to $F$ at fixed $p$, also shown in the insets. Black dashed lines are fits with Eqs.~\eqref{eq:nonLor_peak} and \eqref{eq:largeN_peak}, respectively. $F_\nL$ is shown after filtering out other small peaks. A (blue dashed) Lorentzian curve is included for comparison.}
\label{fig:overview}
\end{figure}

The study of these far-from-equilibrium universality classes in isolated systems has so far primarily focused on the properties of \emph{equal-time} momentum distribution functions, $f(t,p)$.
These functions describe the occupancy of momentum modes $f\sim\langle \hat{a}^\dagger_p \hat{a}_p \rangle$ for a suitably defined basis of excitations $\hat{a}_p$.
The typical scenario is depicted in Fig.~\ref{fig:overview}a. 
Starting with high occupation numbers, which may be obtained, e.g., from instabilities or strong cooling quenches \cite{Prufer:2018hto,Erne:2018gmz}, the system quickly approaches an attractor solution characterized by self-similar scaling, 
$f(t,p) = t^\alpha f_S(t^\beta p)$, also referred to as non-thermal fixed point. During this phase,
the evolution is determined by the universal exponents $\alpha$, $\beta$, and the universal function $f_S$, which are largely insensitive to system parameters and details of the initial conditions.

Scalar field theories with $O(N)$ symmetry 
have been shown to exhibit such universal dynamics. In the \emph{infrared}, they are characterized by $\alpha=\beta d$ and $\beta\approx 1/2$ in $d$ spatial dimensions \cite{PineiroPRD92}. The physics is linked to particle number transport towards low momenta and the growth of a zero-mode condensate. 
Remarkably, previous works have found $\alpha$, $\beta$ and the form of $f_S$ to be universal for all values of $N$ (see Fig.~\ref{fig:overview}b), including both relativistic $O(N)$ and non-relativistic $U(N)$ theories describing ultracold Bose gases~\cite{PineiroPRD92,WalzPRD97,ChantesanaPRA99,GasenzerMikheevPRA99,MoorePRD93}. 
The origin of this universality has remained so far a mystery.
Both exponents and scaling function have been successfully calculated using a large-$N$ kinetic theory, which describes elastic 
collisions of quasiparticles with free dispersion and a renormalized interaction~\cite{PineiroPRD92,Berges:2010ez,WalzPRD97,ChantesanaPRA99,GasenzerMikheevPRA99}. 
However, it is unclear if and why this description should apply at small $N$ as well. At the same time, descriptions based on defects, e.g.~vortices, have provided alternative explanations of related models at small $N$~\cite{NowakPRA85,Karl:2016wko,SchlichtingDeng_PRA97,SextyGasenzerPLB2012}.

In this Letter, we resolve this long-standing puzzle on the universality observed in $O(N)$ scalar theories using instead \emph{unequal-time} (two-point) correlation functions.
By providing information on both occupancies and dispersion relations, these observables allow to identify the dominant far-from-equilibrium excitations~\cite{BoguslavskiLappiPRD98,APO_PRL122},
and further provide access to the universal dynamical exponent $z$~\cite{Schachner:2016frd,MaragaPRE92,ChiocchettaPRL118,Aarts:2001yx,QuantumPTSachdev,Berges:2009jz,Schlichting:2019tbr}. 
As our main result, we find that the previously believed $N$-universality actually breaks up into (at least) two clearly distinct universality classes characterized by different phenomena, which are, however, almost indistinguishable from equal-time correlators and even the dynamical critical exponent $z$ alone.



{\it $O(N)$ theories and statistical function.---}We consider an $O(N)$-symmetric scalar field theory for relativistic scalar fields $\varphi_a(t,\mathbf{x})$, $a=1,\ldots,N$, in $d=3$ spatial dimensions with classical action ($c=k_{\mathrm{B}}=\hbar =1$)
\begin{equation}
S[\varphi] = \int_{t,\mathbf{x}} \left[ \frac{1}{2} \partial^\mu \varphi_a \partial_\mu \varphi_a - \frac{m^2}{2}  \varphi_a \varphi_a - \frac{\lambda}{4! N} \left( \varphi_a \varphi_a \right)^2 \right].
\label{eq_class_action}
\end{equation}
Here, $\int_{t,\mathbf{x}} \equiv \int \mathrm{d} t \int \mathrm{d}^3 x$ and sum over repeated indices is implied. We consider a weak coupling $\lambda\ll1$ and different values of $m$. 
However, since field fluctuations generate an effective mass $\mEff > m$, the exact value of $m$ is not relevant for the infrared physics discussed in this work.

We focus first on the unequal-time 
\emph{statistical function}, defined for a translation invariant system as the anticommutator expectation value of scalar Heisenberg field operators $\hat\varphi_a$
\begin{align}
	F(t,t',\mbf x-\mbf x') =&\, \frac{1}{2N}\big\langle \left\{ \hat\varphi_a(t,\mbf x) , \hat\varphi_a(t',\mbf x') \right\} \big\rangle_c,
\label{eq:Fdef}
\end{align}
where `c' denotes the connected part. 
Introducing the center and relative time coordinates, $\tau\equiv(t+t')/2$ and $\Delta t=t-t'$, we Fourier transform it according to $F(\tau,\omega,\mbf p)\equiv \int d\mbf x \int d\Delta t\; e^{i(\omega\Delta t-\mbf p\mbf x)} F(t,t',\mbf x)$. 

The statistical function can be seen as an unequal-time generalization of the distribution function, which at low momenta $p \lesssim \mEff$ is given by $f(t,\mbf p) \approx F(t,t,\mbf p)\, \mEff$ \footnote{More generally, it is usually defined as $f(t,p)\equiv[F(t,t,\mbf p)\partial_t\partial_{t'} F(t,t',\mbf p)|_{t=t'}]^{1/2}$ \cite{Berges:2004yj}.}. 
$F$ contains not only information about the occupancy of excitations in the system, but also about their frequency dependence. 
Thus, the information contained in $F$ can be crucial to understand \emph{which} excitations dominate the dynamics of a system.

We consider far-from-equilibrium initial conditions with large (Gaussian) fluctuations up to a characteristic scale $Q$ as given by $f(t=0,\mbf p) = \frac{n_0}{\lambda}\, \Theta(Q - p)$. 
Due to the initial `overoccupation' of mode excitations around $Q$, the subsequent redistribution dynamics is dominated by transport of particles to lower momenta, and is characterized by universal scaling in the infrared as explained in the introduction. 

To describe the system we employ classical-statistical simulations (TWA), which are justified in the limit of high occupancies and small couplings as considered here \cite{Aarts:2001yn,Smit:2002yg,Berges:2013lsa,PolkovnikovAP325}.
We perform large-scale simulations averaging over up to $40$ runs using $n_0 = 100$, and give all dimensionful quantities in units of $Q$.
We use either $m=0$ or $m=0.5$ and extract $M$ from our data.
$F$ is computed from a classical correlation function, and
the relative-time Fourier transforms are performed using standard signal-processing methods.
If not stated otherwise, we show data for $\tau = 1000$. 
Further details are given in the Supplementary Material~\cite{Supplemental}.

\begin{figure}[t!]
\centering
\includegraphics[width=\columnwidth]{\pToFigs/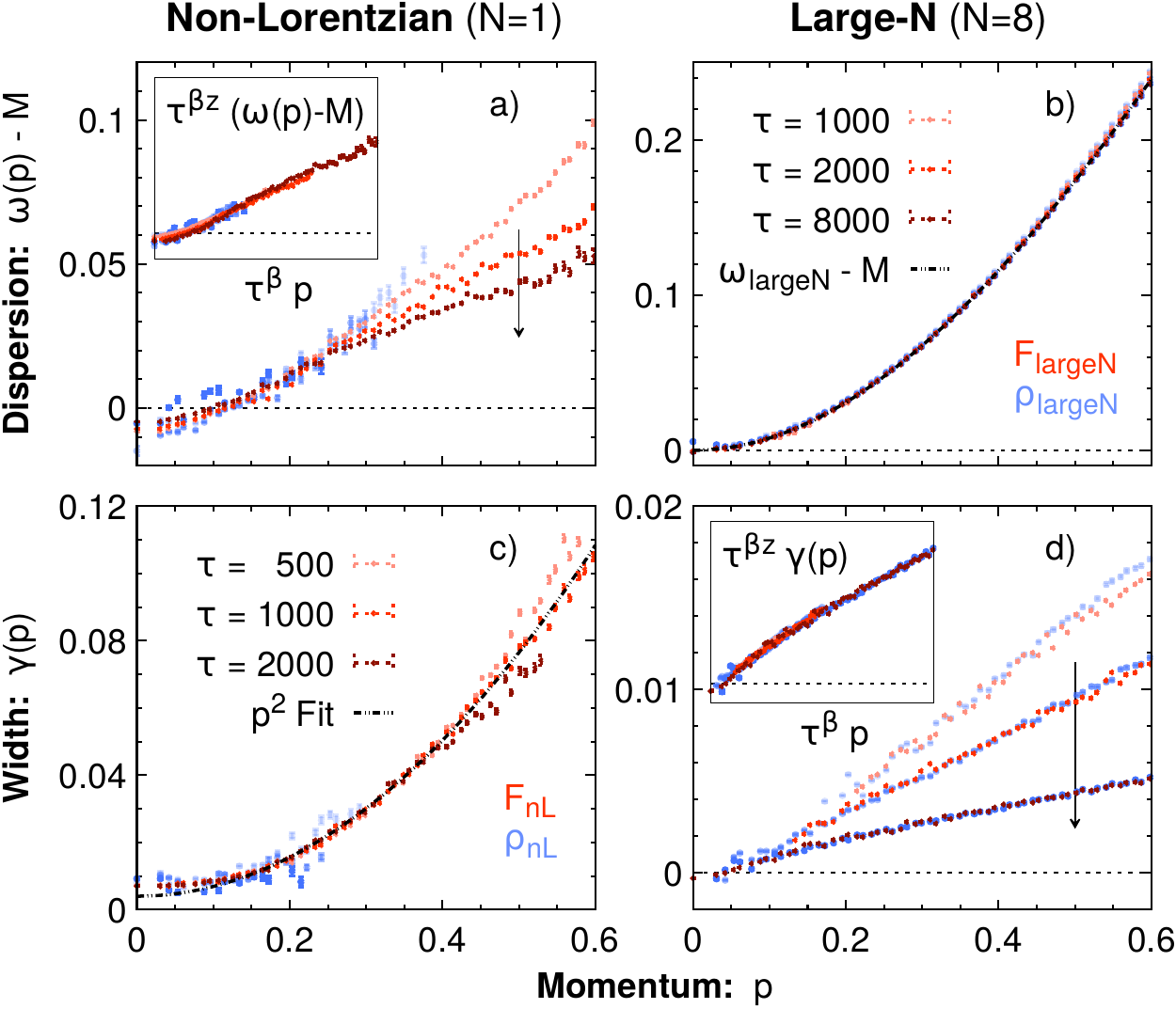}
\caption{Dispersion relations $\omega(p)-\mEff$ (top) and peak width $\gamma(p)$ (bottom) at different times for the dominating peak of $N=1$ (left) and $N=8$ (right), extracted from $F$ (red) and $\rho$ (blue). Arrows mark their time evolution. Insets show rescaled data ($\beta = 1/2$, $z=2$).}
\label{fig:N1N8_dispgamma}
\end{figure}



{\it Large-$N$ peak.---}We consider first the dynamics in the large-$N$ limit.
In general, the statistical function $F$ exhibits several peaks at different frequencies. 
However, for large $N$, we find the signal to be clearly dominated by one single contribution, as shown for $N=8$ in Fig.~\ref{fig:overview}e. We refer to it as the `large-$N$ peak'. 
This peak is depicted for fixed $p$ in the inset (red points), including error bars. It is well-described by a Lorentzian parametrized as
\begin{equation}
	F_\Lor(\tau,\omega,p) \simeq \frac{A_\Lor(\tau,p)\, \gamma_\Lor(\tau,p)}{\left(\omega-\omega_\Lor(p)\right)^2 + \gamma_\Lor(\tau,p)^2},
\label{eq:largeN_peak}
\end{equation}
which is shown as a black dashed line and accurately agrees with the data.

The results of the fitting procedure with Eq.~(\ref{eq:largeN_peak}) are shown in the right column of Fig.~\ref{fig:N1N8_dispgamma} for $N=8$. The dispersion relation (Fig.~\ref{fig:N1N8_dispgamma}b) is time-independent and accurately agrees with a free-particle dispersion of the form
$\omega_\Lor(p)=\sqrt{p^2+\mEff^2}$.
Hence, the dispersion scales quadratically at low momenta as $\omega_\Lor-\mEff \sim p^2/(2\mEff)$, implying a dynamical critical exponent $z=2$.

The decay rate, on the other hand, decreases with time and depends approximately linearly on momentum at low $p$ (Fig.~\ref{fig:N1N8_dispgamma}d). As shown in the inset, its time evolution obeys self-similar scaling given by $\gamma_\Lor(\tau,p)=\tau^{-\beta z}\gamma_{\Lor,S}(\tau^\beta p)$ with $\beta=1/2$ and $z=2$, consistent with the dispersion.
Interestingly, we find that the lifetime of the large-$N$ quasiparticles grows with time since $\gamma_\Lor(\tau,p)\rightarrow0$, such that the form of the large-$N$ peak approaches a $\delta$-function. 

The dominance of this peak implies $F_\Lor(\tau,\omega,p) \approx F(\tau,\omega,p)$ and $A_\Lor(\tau,p)\approx f(\tau,p)/M$. 
Together with the scaling behavior of the dispersion and width, this leads to a self-similar evolution 
\begin{align}
\label{eq:Flor_selfsim_w}
 F(\tau,\omega,p)=\tau^{\alpha+\beta z}\,F_{S}\left(\tau^{\beta z}(\omega-\mEff),\tau^\beta p\right).
\end{align}

To understand the nature of the large-$N$ excitations, we study the behavior of fluctuations with the classical equation of motion $\left[ \partial_\mu \partial^\mu + m^2 + \frac{\lambda}{6N} \varphi_b\varphi_b \right] \varphi_a = 0$ (see Supplemental~\cite{Supplemental} for details).
This equation has stable rotating solutions parametrized by $\vec{\varphi}_0(t,\mbf x)=|\vec{\varphi}_0| \mbf e(t)$, where $\ddot{\mbf e}(t)=-\mEff^2\, \mbf e(t)$. 
They correspond to rotations in a two-dimensional hyperplane in $\varphi$-space, which have been observed numerically~\cite{MoorePRD93}.  
By studying linear fluctuations around these solutions,
we find two in-plane (phase and radial) excitations, and a set of $\sim \!\!\! N$ out-of-plane excitations which correspond to rotations perpendicular to the plane spanned by $\mbf e(t)$ and $\dot{\mbf e}(t)$. The out-of-plane excitations have a $\sqrt{p^2+\mEff^2}$ dispersion and dominate in the large-$N$ limit.
These excitations correspond to the large-$N$ peak discussed here.



{\it $N=1$ non-Lorentzian peak.---}We consider now the statistical function $F$ of a single-component theory as shown in 
Fig.~\ref{fig:overview}d. 
At low momenta $p \lesssim 0.4$
it is again dominated by a single peak \footnote{At larger momenta, a Bogoliubov-like peak with a linear dispersion at low momenta provides the dominant contribution to $F$, visible as the upper branch in the plot. We derive its dispersion in~\cite{Supplemental} and will discuss this and other excitations in a forthcoming work \cite{BogusPinPRDinprep}.}, shown for fixed $p$ in the inset (red points).
While a Lorentzian fit (blue dashed line) with (\ref{eq:largeN_peak}) fails to capture the tails of the peak, we find it to be phenomenologically well described by (black dashed curve)
\begin{equation}
	F_\nL(\tau,\omega,p) \simeq \frac{\pi}{2}\,\frac{A_\nL(\tau,p)}{\gamma_\nL(p)} \sech \left[ \frac{\pi}{2}\,\frac{\omega-\omega_\nL(\tau,p)}{\gamma_\nL(p)} \right],
\label{eq:nonLor_peak}
\end{equation}
where the subscript `$\nL$' stands for non-Lorentzian. 

We employ this form as a fit function to extract the properties of this peak 
leading to the results in the left column of Fig.~\ref{fig:N1N8_dispgamma}. 
The dispersion relation $\omega_\nL(\tau,p)$ (Fig.~\ref{fig:N1N8_dispgamma}a) 
is approximately linear and
obeys a self-similar scaling form $\omega_\nL(\tau,p)-\mEff=\tau^{-\beta z}\tilde{\omega}_S(\tau^\beta p)$ with exponents $\beta = 1/2$ and $z = 2$, as shown in the inset. The decay rate $\gamma_\nL(p)$ (Fig.~\ref{fig:N1N8_dispgamma}c) is found to be instead time-independent and to scale as $\gamma_\nL\sim p^{z}$. This implies a self-similar evolution
as in Eq.~(\ref{eq:Flor_selfsim_w}), notably with the same scaling exponents as for the large-$N$ peak and $A_\nL(\tau,p) \approx f(\tau,p)/M$.

Remarkably, we find the properties of this non-Lorentzian peak to be identical to the infrared peak found in Ref.~\cite{APO_PRL122} for a non-relativistic $U(1)$ complex scalar theory.
This can be explained by noticing that at small momenta, $p\ll \mEff$, particle number changing processes are suppressed and the $O(1)$ theory is hence described by an emergent non-relativistic $U(1)$ theory.
The mapping can be made more rigorous by defining the non-relativistic degrees of freedom $\psi=e^{iMt}[\sqrt{\omega_x}\,\varphi + i/\sqrt{\omega_x}\,\pi]/\sqrt{2}$ with $\pi=\dot{\varphi}$ and $\omega_x=\sqrt{\mEff^2-\nabla^2}$~\cite{Namjoo:2017nia,SchlichtingDeng_PRA97}.

\begin{figure}[t!]
\centering
\includegraphics[width=\columnwidth]{\pToFigs/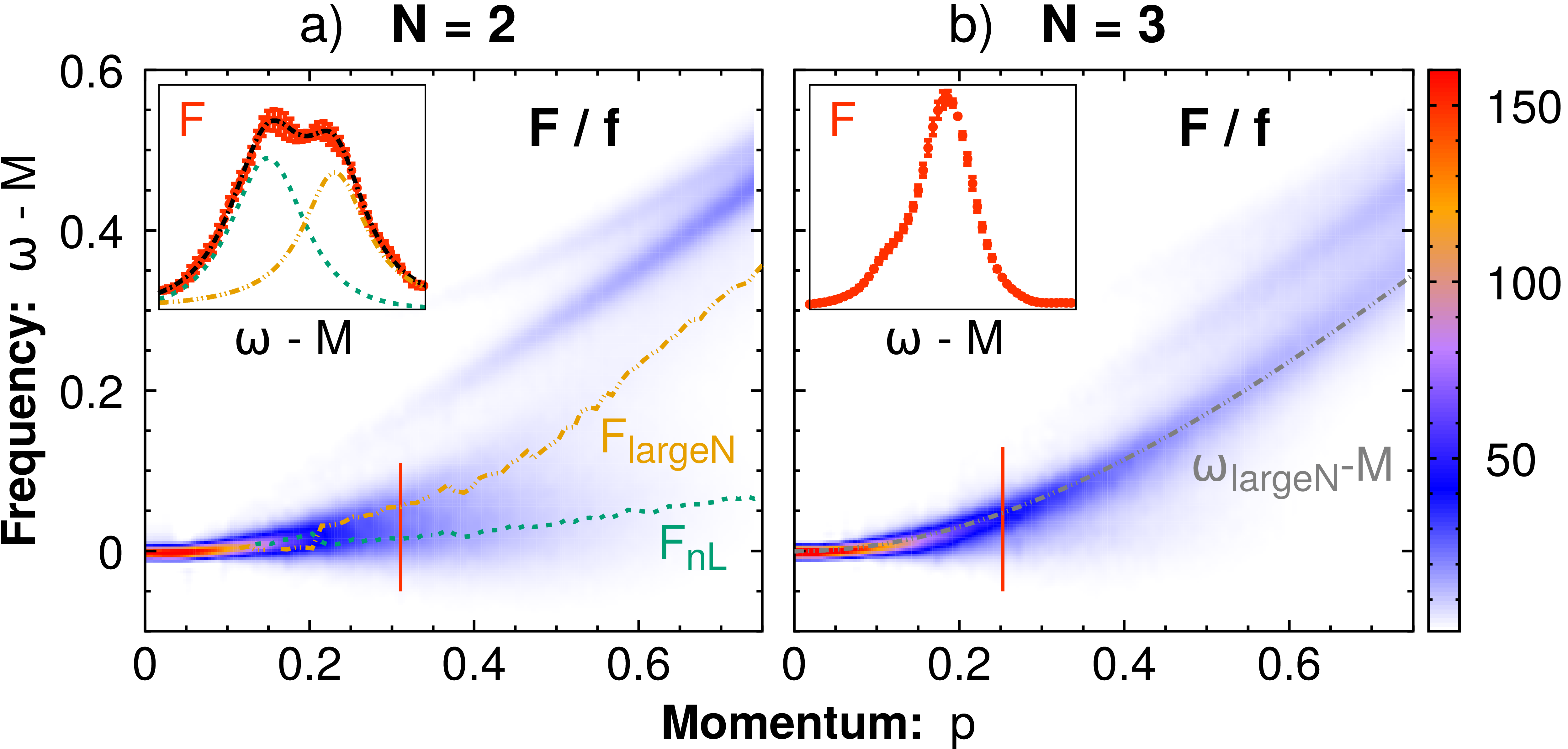}
\caption{Statistical function $F(\tau,\omega,p)/f(\tau,p)$ for $N=2$ (left, a) and $N=3$ (right, b). The insets show $F$ (red) for the fixed momentum marked by the vertical red lines. \emph{Left:} green/yellow dashed lines show the peak positions extracted from fits, as well as the fit results in the inset. \emph{Right:} a relativistic dispersion (gray) is added for comparison.}
\label{fig:intermediateN}
\end{figure}



{\it Intermediate $N$.---}Which of these two distinct physical phenomena, if any, dominates for intermediate $N$ at low momenta is a priori not obvious.

We find that the $N=2$ theory is a special case.
The statistical function $F$ appears to have two distinct contributions with a similar weight at low momenta, as shown in 
Fig.~\ref{fig:intermediateN}a. The inset shows a fit (black dashed line) to these peaks with the sum of both functions (\ref{eq:nonLor_peak}) and (\ref{eq:largeN_peak}). 
Each of these peaks is included in the inset as separate dashed lines (green and yellow), and their  respective dispersions are shown in the main plot. We find that the dispersion of the left peak agrees with $\omega_\nL$ of the non-Lorentzian peak in $O(1)$ theory, while the dispersion of the right peak obeys approximately $\omega_\Lor$ of the large-$N$ peak.
However, at low momenta both contributions overlap so strongly that they become almost indistinguishable. Therefore, their $p\rightarrow 0$ asymptotic behavior is hard to extract.

For $N\geq3$ we find the infrared dynamics to be dominated by the large-$N$ peak. The statistical function for $N=3$ is shown in 
Fig.~\ref{fig:intermediateN}b and for fixed momentum in the inset. The dominant peak has a dispersion that agrees with $\omega_\Lor(p)$ (gray dashed line), and we confirmed that the width shows a similar behavior as in the large-$N$ limit. We note, however, that for small $N$ we find evidence of an additional contribution overlapping with the main peak at lower frequencies. 
Based on the above, this is possibly related to a non-Lorentzian contribution, which appears to quickly disappear as $N$ or momentum increases.



\begin{figure}[t!]
\centering
\includegraphics[width=\columnwidth]{\pToFigs/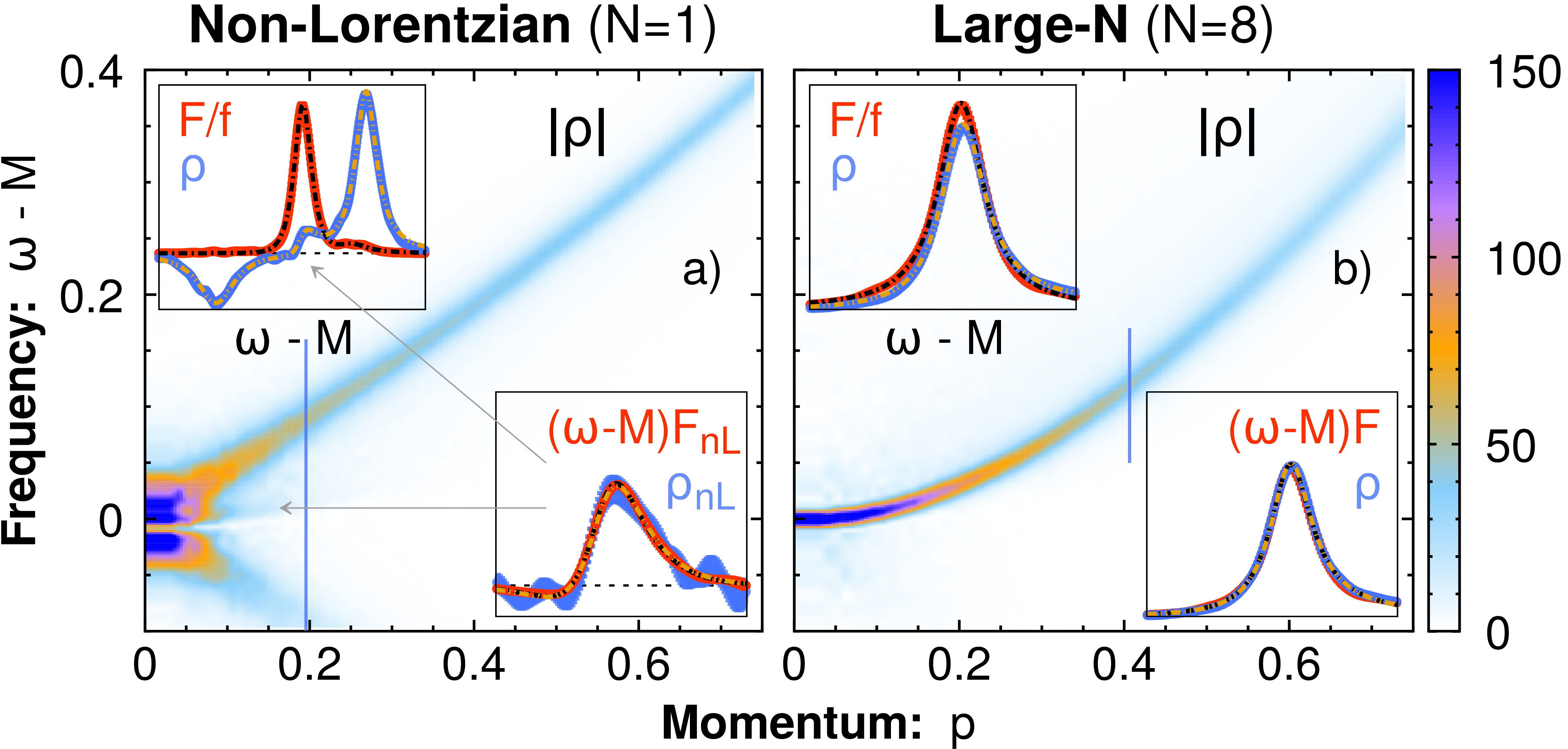}
\caption{Spectral function $|\rho|$ for $N=1$ (left, a) and $N=8$ (right, b). The upper insets show $F/f(\tau,p)$ and $\rho$ for fixed $p$ marked by vertical blue lines. The lower insets show the validity of Eq.~(\ref{eq:genT_FDR}) by comparing $(\omega-M)F$ and $\rho$, both normalized to weight one. The ratio of normalizations gives $T_\nL(\tau,p)$. Black/yellow dashed lines are fits to $F$, $\rho$.}
\label{fig:FDR}
\end{figure}

{\it Spectral function and generalized FDR.---}To further characterize the non-Lorentzian and large-$N$ Lorentzian peaks, we study the spectral function defined as
\begin{align}
	\rho(t,t',\mbf x-\mbf x') =&\, \frac{i}{N}\big\langle \left[ \hat\varphi_a(t,\mbf x) , \hat\varphi_a(t',\mbf x') \right] \big\rangle.
\label{eq:rhodef}
\end{align}
At equal times, $t=t'$, it is determined by the equal-time commutation relations, $\rho|_{t=t'}=0$ and $\partial_t\rho|_{t=t'}=\delta(\mbf x-\mbf x')$. For unequal times, this quantity encapsulates the linear response of the system to perturbations and thus contains information about the low-lying excitations of the system. To compute it we employ a linear response approach as described in Refs.~\cite{BoguslavskiLappiPRD98,APO_PRL122} (see~\cite{Supplemental} for details).

In Fig.~\ref{fig:FDR} we show color plots of the spectral function for $N=1$ and $N=8$. 
In general, we find that $\rho$ shows the same peak structure as $F$ but with different relative weights between the peaks. However, since $\rho$ does not contain information about occupancies, the weight of the peaks does not reveal the dominant excitations.

In particular, for $N=1$ we find that the non-Lorentzian peak in $\rho$ has a very small weight \footnote{This is probably why it was not observed for the $U(1)$ theory in Ref.~\cite{APO_PRL122}, which had a smaller frequency resolution.} (upper inset of Fig.~\ref{fig:FDR}a), which becomes visible only at low momenta.
Nevertheless, this peak has the same dispersion and width as the peak in $F$ (blue points in Figs.~\ref{fig:N1N8_dispgamma}a,c). In fact, we find that the non-Lorentzian peaks in $F$ and $\rho$ fulfill a generalized fluctuation-dissipation relation (FDR) given by
\begin{align}
	F_\nL(\tau,\omega,p) \simeq&\, \frac{T_\nL(\tau,p)}{\omega-\mu}\, \rho_\nL(\tau,\omega,p).
\label{eq:genT_FDR}
\end{align}
We show this in the lower inset of Fig.~\ref{fig:FDR}a by filtering out all peaks but the non-Lorentzian peak~\cite{Supplemental}.
Here, $\mu \equiv \mEff$ is an effective chemical potential linked to the approximate conservation of particle number at low momenta.
Equation (\ref{eq:genT_FDR}) is reminiscent of the equilibrium FDR \cite{KadanoffBaym},
except with a mode-dependent temperature $T_\nL(\tau,p)$. 

In the large-$N$ limit, the spectral function is dominated by the large-$N$ peak (Fig.~\ref{fig:FDR}b) with a dispersion and width which also match the results for the corresponding large-$N$ peak in $F$ (Figs.~\ref{fig:N1N8_dispgamma}b,d). The two peaks can again be related through a generalized FDR as in Eq.~(\ref{eq:genT_FDR}) 
(lower inset of Fig.~\ref{fig:FDR}b). However, since the large-$N$ peak dominates in both $F$ and $\rho$, and since its width becomes narrower with time, Eq.~(\ref{eq:genT_FDR}) can be simplified in the long-time, large-$N$ limit to
\begin{align}
	F_\Lor(\tau,\omega,p) \simeq&\, f(\tau,p)\, \rho_\Lor(\tau,\omega,p).
\label{eq:genkin_FDR}
\end{align}
This is again similar to the thermal case, except with a time-dependent non-equilibrium distribution, and is reminiscent of kinetic approximations~\cite{Berges:2004yj}. 
At intermediate $N\geq3$ we find small deviations from this behavior, which vanish as $N\rightarrow\infty$.



{\it Discussion.---}Our results reveal the existence of (at least) two distinct infrared universality classes governed by two different types of phenomena, despite being characterized by very similar universal exponents. 

For $N\geq3$, the dynamics is dominated by a Lorentzian large-$N$ peak.
It results from a set of excitations with relativistic dispersion $\sqrt{p^2+\mEff^2}$ which dominate 
due to their number scaling with $N$. The fact that this physics dominates even at low $N$ is, however, remarkable, especially for $N=3$ where only one such excitation exists.
A possible explanation for this is the fact that the dispersion $\omega_p-M$ scales as $\sim p^2$ in the infrared. Thus, at low momenta large-$N$ excitations are energetically easier to excite compared to the Bogoliubov mode which scales as $\sim p$, and other excitations with larger effective mass~\cite{Supplemental}.
Apart from this, given that the lifetime of these quasiparticles grows with time, and the fact that they fulfill the generalized FDR of Eq.~(\ref{eq:genkin_FDR}), our results validate the analytic large-$N$ kinetic theory used in Refs.~\cite{PineiroPRD92,ChantesanaPRA99,WalzPRD97}. 

For $N=1$, the dynamics is instead dominated by a peak of non-Lorentzian shape with time-dependent dispersion and time-independent quadratic width, which coincides with the findings for the non-relativistic $U(1)$ theory~\cite{APO_PRL122}. This hints to a common origin for these infrared excitations.
We suggest that this non-Lorentzian peak in $O(1)$ corresponds to vortex line excitations in effective non-relativistic degrees of freedom.
This claim is motivated by previous works on $U(1)$ dynamics~\cite{NowakPRB84,NowakPRA85}, and $O(1)$ dynamics~\cite{MoorePRD93,SchlichtingDeng_PRA97}, where evidence of vortex excitations was seen in real-space snapshots of the field.

Our results for $N=2$ show a mixture of two different contributions. We found evidence that one contribution has 
a $\sqrt{p^2+\mEff^2}$ dispersion, analogously to the large-$N$ peak.
Analytically, however, we do not find any excitation with such a dispersion for the $O(2)$ theory~\cite{Supplemental}.
The second contribution was found to share the same properties as the non-Lorentzian peak of $O(1)$. Thus, we hypothesize that these two peaks possibly originate from vortex-type excitations and domain walls, such as those observed in Refs.~\cite{SextyGasenzerPLB2012,MoorePRD93}.
In turn, this could explain why the non-Lorentzian contribution is suppressed at large $N$. Vortex and domain wall excitations can be easily unwinded or smoothed out in configuration space when $\varphi_a$ has $N\geq3$ components. Thus, they are not stable enough to contribute to the self-similar dynamics at large $N$.



{\it Conclusion.---}In this work, we have disentangled the physical origin of the infrared universal dynamics of $O(N)$ scalar theories.
Despite equal-time properties being universal for all $N$, unequal-time correlators have allowed us to identify at least two distinct universality classes as a function of $N$:
non-Lorentzian excitations for $N=1$ and Lorentzian rotational excitations for $N\geq3$, while for $N=2$ we find a mixture. This constitutes a crucial step in classifying universality classes far from equilibrium. 
In a broader context, our work shows the importance of the unequal-time statistical function $F$ to reveal the dominant physical phenomena in far-from-equilibrium systems,
and will potentially trigger future research into this direction.
In particular, measuring unequal-time functions~\cite{KastnerPRA96,UhrichQST2019} could substantially improve our understanding of running cold-atom experiments.

\begin{acknowledgments}
{\it Acknowledgements.---}We are grateful to J.~Berges, T.~Gasenzer, T.~Lappi and S.~Schlichting for helpful discussions and collaboration on related work. The authors wish to acknowledge CSC – IT Center for Science, Finland, and the Vienna Scientific Cluster (VSC) for computational resources. 
\end{acknowledgments}

\bibliography{ON_spectral_bibliography}


\vfill
\newpage

\onecolumngrid
\vspace{\columnsep}
\begin{center}
\textbf{\large Supplementary Material: \\ Unraveling the nature of universal dynamics in $O(N)$ theories}
\end{center}
\vspace{\columnsep}
\twocolumngrid

\setcounter{equation}{0}
\setcounter{figure}{0}
\setcounter{table}{0}
\setcounter{page}{1}
\makeatletter
\renewcommand{\theequation}{S\arabic{equation}}
\renewcommand{\thefigure}{S\arabic{figure}}
\renewcommand{\bibnumfmt}[1]{[#1]}
\renewcommand{\citenumfont}[1]{#1}


\section{Technical implementation}

Here we provide more details on our simulations. As explained in the main text, we consider $O(N)$-symmetric relativistic scalar theories given by the action
$S[\varphi] = \int_{t,\mathbf{x}} \left[ \frac{1}{2} \partial^\mu \varphi_a \partial_\mu \varphi_a - \frac{m^2}{2}  \varphi_a \varphi_a - \frac{\lambda}{4! N} \left( \varphi_a \varphi_a \right)^2 \right]$.
Because of the large occupation numbers at low momenta in our initial conditions 
\begin{align}
    f(t=0,\mbf p) = \frac{n_0}{\lambda}\, \Theta(Q - p),
    \label{eq:IC_box}
\end{align}
quantum vacuum contributions are suppressed by a factor of $\lambda$ relative to $f(0,\mbf p)$. In the weak-coupling limit $\lambda \rightarrow 0$ with $\lambda f(t,p)$ held fixed, the original quantum field theory is accurately mapped onto a classical-statistical field description \cite{Aarts:2001yn,Smit:2002yg,Berges:2013lsa}. In particular, the dynamics is not governed by the lattice cutoff but by physical scales like $Q$ instead. 
We ensured that our results are insensitive to descretization parameters like the lattice spacing or the volume by varying them.

In the numerical approach, classical fields $\varphi_a(t,\mbf x)$ and their conjugate momenta $\pi = \partial_t \varphi$ are discretized on a lattice with lattice spacing $a_s$ and volume $V = (N_s a_s)^3$. Unless stated otherwise, we used $N_s = 256$ and $a_s = 0.8$ in units of $Q$. 
The above initial conditions \eqref{eq:IC_box} are implemented by sampling the field as 
\begin{align}
    \varphi_a(0,\mbf p)&=\sqrt{f(0,p)/p}\;c_a(\mbf p), \nonumber \\
    \pi_a(0,\mbf p)&=\sqrt{p f(0,p)}\;\tilde{c}_a(\mbf p),
\end{align}
with independent Gaussian random numbers fulfilling $\left\langle c_{a}(\mbf p) \left(c_{b}(\mbf q)\right)^* \right\rangle_{\cl} = V \delta_{\mbf p, \mbf q} \delta_{a,b}$ and similarly for $\tilde{c}$. The fields are evolved by solving the classical Hamilton equations of motion, where we use $a_t = 0.05 a_s$ for the time step. To reduce lattice artefacts, we use a fourth order discretization scheme for second order spatial derivatives, as described in Ref.~\cite{Micha:2002ey}. 

In the framework of classical-statistical simulations, the statistical function $F$ can be straightforwardly computed as \cite{Berges:2004yj}
\begin{align}
    F(t,t',p)= \frac{1}{VN} \left\langle \varphi_a(t,\mbf p) \varphi_a(t',-\mbf p) \right\rangle_\cl
\end{align}
where $\langle\cdot\rangle_\cl$ denotes average over initial conditions and over the direction of $\mbf p$. 

To compute the spectral function we employ a linear response approach similar to Refs.~\cite{BoguslavskiLappiPRD98,APO_PRL122}. In essence, we perturb the system as $\varphi_a\rightarrow \varphi_a+\delta\varphi_a$ with a source $j_a(t,\mbf x)=j^0_a(\mbf x) \delta(t-t')$ at time $t'$ and time evolve the perturbation $\delta\varphi_a$ according to the linearized equations of motion $\partial_{t}\delta\varphi_a = \delta\pi_a$ and 
\begin{align}
 \label{eq_linflucts_eom}
 \partial_{t} \delta\pi_a =& \Delta\, \delta\varphi_a - \left(m^2 + \frac{\lambda}{6 N} \varphi_b \varphi_b\right) \delta \varphi_a \nonumber \\
 &~~ + \frac{\lambda}{3 N} \varphi_a \varphi_b\, \delta \varphi_b - j^0_a\, \delta(t - t')\,.
\end{align}
We choose to perturb with random Gaussian source fields fulfilling $\langle j^0_a(\mbf p) (j^0_b(\mbf q))^* \rangle_j = V \delta_{\mbf p,\mbf q}\delta_{ab}$, where $\langle\cdot\rangle_j$ denotes average over the perturbations. The retarded part of the spectral function results from linear response theory as
\begin{align}
    \theta(t-t')\rho(t,t',p)=\frac{1}{V\,N}\langle \langle \delta\varphi_a(t,\mbf p) \rangle_\cl (j^0_a(\mbf p))^* \rangle_j\,.
\end{align}
Our choice of $j^0_a(\mbf p)$ ensures that the canonical commutation relations $\lim_{t \rightarrow t'} \rho(t, t', p) = 0$ and $\lim_{t \rightarrow t'} \partial_t\rho(t, t', p) = 1$ are implemented correctly. 

After Fourier transforming with respect to relative time $\Delta t=t-t'$, the correlations become
\begin{align}
    F(\tau,\omega,p)&\equiv \int d\Delta t\; e^{i\omega\Delta t} F(t,t',p) \nonumber \\
    &= 2\int_0^{\infty} d\Delta t\; \cos(\omega\Delta t)\, F(t,t',p) \\
    \rho(\tau,\omega,p)&\equiv -i\int d\Delta t\; e^{i\omega\Delta t} \rho(t,t',p) \nonumber \\
    &= 2\int_0^{\infty} d\Delta t\; \sin(\omega\Delta t)\, \rho(t,t',p)\,
\end{align}
with central time $\tau\equiv(t+t')/2$.
We used that $F$ ($\rho$) is even (odd) in $\Delta t$, such that their Fourier transforms are real-valued functions of $\omega$. Strictly speaking, these symmetries only hold for fixed $\tau$. 
Because of the slow $\tau$-dependence of our observables and for the considered time windows of $\Delta t$, we find the transforms at fixed $t'$ to accurately approximate the ones at fixed $\tau$. 
In practice, we hold $t'$ fixed and compute the transforms as 
\begin{align}
 \label{eq:Ftrafo_pract}
    F(\tau,\omega,p) &\approx 2\!\!\int_{0}^{\Delta t_{\text{max}}} d\Delta t\; \cos(\omega\Delta t)\,h(\Delta t)\, F(t,t',p) \\
\label{eq:rhotrafo_pract}
    \rho(\tau,\omega,p) &\approx 2\!\!\int_{0}^{\Delta t_{\text{max}}} d\Delta t\; \sin(\omega\Delta t)\,h(\Delta t)\, \rho(t,t',p),
\end{align}
where we typically employ $\Delta t_{\text{max}} \approx 300 - 500$. To smoothen the resulting curves, we use standard signal processing techniques by employing the Hann window function 
\begin{align}
    h(\Delta t) = \frac{1}{2} \left(1 + \cos\frac{\pi \Delta t}{\Delta t_{\text{max}}} \right) ,
\end{align}
and zero-padding. The latter effectively amounts to evaluating \eqref{eq:Ftrafo_pract} and \eqref{eq:rhotrafo_pract} at more intermediate frequencies than provided by the usual discrete Fourier transform. We checked that these techniques do not alter the peak structure nor the peak forms considerably, but mainly reduce background ringing. 

Correlation functions at fixed momentum $p$ are shown in the main text with error bars. To obtain the statistical errors, we first Fourier transform our data for each run and subsequently average over the results. 

Because of the effective mass $\mEff > m$, generated by fluctuations, the exact value of $m$ is not relevant for the infrared physics discussed in this work. In particular, the correlation functions $F$ and $\rho$ plotted as functions of $\omega - \mEff$ are independent of $m$ at low momenta $p < \mEff$.
For the $N \geq 2$ data shown we used $m=0$, but we have also checked that $m>0$ does not change the results. For $N=1$, the effective mass $\mEff$ decreases as a power law in time for $m=0$, whereas it stays approximately constant for sufficiently large nonzero $m$ \cite{Micha:2002ey,Berges:2013lsa,MoorePRD93}. To avoid this issue, we use $m=0.5$ in all figures with $N=1$ shown in the main text. However, we have checked that simulations with $m=0$ provide the same peak structure and the same properties of the non-Lorentzian peak as for $m > 0$.


\section{Data analysis}

To extract properties of the peaks, we perform fits to the spectrum with different fit forms. We distinguish between Lorentzian and non-Lorentzian peaks, where we use a hyperbolic secant form for the latter. The fit forms for these peaks are 
\begin{align}
 L(\omega; \omega_0,\gamma) \,&= \frac{1}{\gamma}\frac{1}{1 + \left(\frac{\omega-\omega_0}{\gamma} \right)^2} \\
 S(\omega; \omega_0,\gamma) \,&= \frac{\pi}{2\gamma}\, \rm{sech}\!\left( \frac{\pi}{2}\,\frac{\omega-\omega_0}{\gamma} \right) \,.
 \label{eq:sech_fit}
\end{align}
We employ these functions to fit the positive frequency peaks ($\omega_0>0$) of $F$ and $\rho$. Due to symmetries, the same peaks appear as well at negative frequencies $\omega_0 \mapsto -\omega_0$.
However, these negative-frequency contributions are irrelevant for our fits since we have $\gamma/\omega_0 \ll 1$ for all peaks and thus, it is sufficient to perform fits in positive-frequency space.

The corresponding fit functions in $\Delta t$ are 
\begin{align}
 \tilde{L}(\Delta t; \omega_0,\gamma) \,&= \cos\left(\omega_0 \Delta t\right)\, e^{-\gamma |\Delta t|}, \\
 \tilde{S}(\Delta t; \omega_0,\gamma) \,&= \cos\left(\omega_0 \Delta t\right)\, \rm{sech}\!\left(\gamma \Delta t \right) .
\end{align}

In general, we find that fits in $\omega$ are easier to perform and more likely to converge than fits in $\Delta t$. However, the latter are useful when the peak width is very narrow, as is the case for the large-$N$ peak in $N=8$ (Fig.~\ref{fig:N1N8_dispgamma}d in main text). To extract the properties of such narrow peaks, we first performed fits in $\omega$ space with $L(\omega; \omega_0,\gamma)$ and then used the results as initial values for fits in $\Delta t$ with $\tilde{L}(\Delta t; \omega_0,\gamma)$. 

The properties of the dominating peak for $N=1$ were obtained in frequency space. For $F$, we use \eqref{eq:sech_fit} to fit $F$ for $p < 0.15$. At higher momenta, other peaks start to become visible, as one finds in Fig.~\ref{fig:overview}d of the main text. The main additional peaks correspond to a relativistic generalization of Bogoliubov excitations (see next section) with a Lorentzian shape.
While additional peaks will be studied in detail in Ref.~\cite{BogusPinPRDinprep}, the fit function needs to include them at larger momenta, because peaks start to overlap. Hence, to fit $F$ for $p \geq 0.15$ we use
\begin{align}
    &F_1 L(\omega; \omega_1,\gamma_1) + F_\nL S(\omega; \omega_\nL,\gamma_\nL) \nonumber \\
    + \,&F_2 L(\omega; \omega_2,\gamma_2)\,,
    \label{eq:N1_F_fit}
\end{align}
with the fit parameters $\omega_i$, $\gamma_i$ and $F_i$. Here $F_i$ is the weight of each peak, and the sum of all weights equals $F(t,t,p)$. However, since $S$ dominates at low momenta one has $F_\nL \approx F(t,t,p)$. The insets in Figs.~\ref{fig:overview}d and \ref{fig:FDR}a (bottom inset) of the main text show the non-Lorentzian part of $F(\tau,\omega,p)$ (denoted in the plots as $F_\nL$). We obtain this after subtracting the fit results of the Bogoliubov peaks [$F_1$ and $F_2$ peaks in (\ref{eq:N1_F_fit})] from the original signal. 

For the spectral function $\rho$, Bogoliubov excitations are always important for $N=1$, as can be seen in Fig.~\ref{fig:FDR}a of the main text. We still find the same non-Lorentzian peak at low momenta, which appears, however, extremely small and hard to detect without signal processing techniques. To extract its properties, we performed a similar fit as above with
\begin{align}
    &A_1 L(\omega; \omega_1,\gamma_1) + (\omega-\mu) A_\nL S(\omega; \omega_\nL,\gamma_\nL) \nonumber \\
    + \,&A_2 L(\omega; \omega_2,\gamma_2)\,,
    \label{eq:N1_Rho_fit}
\end{align}
where the fit parameters are $\omega_i$, $\gamma_i$, $A_i$ and $\mu$ that turns out to be $\mu \approx \mEff$.
The corresponding results are shown in the left panel of Fig.~\ref{fig:N1N8_dispgamma} of the main text as blue points.

Bogoliubov excitations are also present for other values of $N$. The corresponding peaks generally appear in two branches with dispersions $M\pm\omega_{\text{Bog}}(p)$ symmetric around $M$.
Therefore, we use the mean value of the two dispersions to calculate $\mEff$. For $N=8$, these excitations are barely visible and we use instead the relativistic dispersion $\sqrt{\mEff^2 + p^2}$ to obtain $\mEff$. We checked that both methods, where applicable, provide a consistent value for $\mEff$.


\section{$O(N)$ excitations for $N\geq2$}

We compute in this section the dispersion of the large-$N$ excitations discussed in the main text. For this we consider the $O(N)$ scalar theory defined in Eq.~\eqref{eq_class_action} of the main text for $N\geq2$. The corresponding classical equation of motion reads
\begin{align}
	\left[ \partial_t^2 - \nabla^2 + m^2 + \frac{\lambda}{6N}(\varphi_a\varphi_a) \right] \varphi_b = 0.
\label{eq:clasEOM}
\end{align}
This equation supports homogeneous solutions of the form
\begin{equation}
	\vec\varphi_0(t) = \sqrt{\rho_0}\; \vec e(t),
\label{eq:phi0_sol_largeN}
\end{equation}
where $\rho_0>0$, $\vec \varphi_0$ is an $N$-component vector in $\varphi$-space, and the unit vector $\vec e(t)$ fulfils
\begin{equation}
	\frac{d^2}{dt^2}\, \vec e(t) = - \mEff^2\, \vec e(t).
\label{eq:eom_unitvector}
\end{equation}
The effective mass is given by
\begin{equation}
	\mEff = \sqrt{ m^2 + \frac{\lambda}{6N} \rho_0 }\,.
\end{equation}
Solutions to Eq.~(\ref{eq:eom_unitvector}) can be easily constructed as
\begin{align}
	\vec e(t) = R\, \big[ \cos(\omega_0t) \ket{1} + \sin(\omega_0t) \ket{2} \big],
\label{eq:e_sol}
\end{align}
where $\omega_0\equiv\pm \mEff$, $\ket{1}=(1,0,0,\ldots)^T$, $\ket{2}=(0,1,0,\ldots)^T$, and $R\in O(N)$ is an arbitrary orthogonal matrix, $R R^T = \mathbbm{1}$. These solutions correspond to rotations on an arbitrary but fixed two-dimensional hyperplane in $\varphi$-space that passes through the origin.

In the following, we study the nature of excitations around solutions of the type (\ref{eq:phi0_sol_largeN}). For this we add fluctuations as
\begin{align}
	\vec\varphi(t,\mbf x) =&\, \sqrt{\rho_0 + \delta\rho(t,\mbf x)}\, e^{\Lambda_{ab}\theta_{ab}(t,\mbf x)}\, \vec e(t).
\end{align}
Here, it is assumed that $\delta\rho\ll\rho_0$ and $|\theta_{ab}|\ll1$. The matrices $\Lambda_{ab}$ correspond to the generators of $O(N)$ rotations.
Expanding this expression to linear order in the fluctuations we obtain $\vec\varphi(t,\mbf x)=\vec\varphi_0(t) + \delta\vec\varphi(t,\mbf x)$ with
\begin{align}
	\delta\vec\varphi(t,\mbf x) =&\, \left[ \frac{\delta\rho(t,\mbf x)}{2\sqrt{\rho_0}} + \sqrt{\rho_0 }\, \Lambda_{ab}\,\theta_{ab}(t,\mbf x)\right] \vec{e}(t).
\end{align}

Inserting this into the equation of motion (\ref{eq:clasEOM}) and expanding to linear order leads to the equation
\begin{align}
	&\,2 \left[ \delta\dot\rho + 2\rho_0\,\Lambda_{ab}\,\dot\theta_{ab} \right] \dot{\vec e} \nonumber\\
	&\,+ \left[ \delta\ddot\rho - \nabla^2\delta\rho + \frac{\lambda}{3N}\rho_0\,\delta\rho + 2\rho_0\,\Lambda_{ab} (\ddot\theta_{ab} - \nabla^2\theta_{ab}) \right] \vec{e} = 0.
\label{eq:largeNflucteq1}
\end{align}
Next, we assume without loss of generality 
that $R=\mathbbm{1}$ in Eq.~(\ref{eq:e_sol}), and choose a representation for the generators as
\begin{equation}
	\Lambda_{ab} = \ket{a}\bra{b} - \ket{b}\bra{a}.
\label{eq:ONgenerators}
\end{equation}
These correspond essentially to Pauli $i\sigma_y$ rotations in the plane defined by the axes $a$, $b$.
Using Eq.~(\ref{eq:ONgenerators}), the action of the generators on $\vec{e}$ and $\dot{\vec{e}}$ can be straightforwardly computed.
In this way, we obtain from Eq.~(\ref{eq:largeNflucteq1}) a sum of terms proportional to the vectors $\vec{e}$, $\dot{\vec{e}}$, and $\ket{b}$ ($b\geq3$), which are all orthonormal to each other. The resulting linearized equation of motion can then be projected onto the different axes defined by these vectors, leading to a total of $N$ equations, which we analyze in the following.


\subsection{In-plane excitations}

We start by projecting Eq.~(\ref{eq:largeNflucteq1}) onto $\vec{e}$ and $\dot{\vec{e}}$. This leads to a system of equations for $\delta\rho$ and $\theta_{12}$ given by
\begin{align}
\begin{aligned}
	\left(\partial_t^2 - \nabla^2 + \frac{\lambda}{3N}\rho_0 \right) \delta\rho + 4\rho_0 \omega_0\, \partial_t \theta_{12} =&\, 0,\\
	\left( \partial_t^2 - \nabla^2 \right) \theta_{12} - \frac{\omega_0}{\rho_0} \partial_t \delta\rho =&\,0.
\end{aligned}
\label{eq:N2_inplane_eqs}
\end{align}
The variables $\delta\rho$ and $\theta_{12}$ correspond to radial and phase excitations in the plane spanned by $\vec \varphi_0(t)$ and $\partial_t \vec \varphi_0(t)$.
To solve this, we Fourier transform into frequency $\tilde\omega$ and momentum $\mbf p$ space. Solving for $\delta\rho$ and inserting it into the other equation leads to
\begin{align}
	\left[ \tilde\omega^4 - \left( 2p^2 + \frac{\lambda}{3N}\rho_0 + 4\mEff^2 \right)\tilde\omega^2 \right.&\, \nonumber\\
	\left.+ p^2\left( p^2 + \frac{\lambda}{3N}\rho_0 \right) \right]&\, \theta_{12} = 0.
\end{align}
The four solutions to this are given by $\tilde\omega_{\pm_1\pm_2} \equiv \pm_1 \tilde\omega_{\pm_2}$, where $\pm_1$ and $\pm_2$ are independent, and
\begin{align}
	\tilde\omega_{\pm}=\sqrt{ p^2 + 2\mEff \left(\mEff + g\rho_0 \pm \sqrt{p^2 + (\mEff + g\rho_0)^2} \right) }.
\end{align}
Here, we defined the effective interaction coefficient
\begin{equation}
	g\equiv\frac{\lambda}{12N\mEff}.
\label{eq:g_largeN}
\end{equation}
These excitations correspond to a Bogoliubov-type phase excitation $\omega_\text{Bog}(p)\equiv \tilde\omega_-$ which becomes linear in $p$ at low momentum, and a massive radial excitation $\omega_\text{rad}(p) \equiv \tilde\omega_+$ with $\omega_\text{rad}(p=0)=2\sqrt{M(M+g\rho_0)}$.

Note that in numerical simulations we look at correlation functions in the original $\varphi$ basis, instead of in the phase and radial variables. Because of this, the dispersions numerically presented in the main text for $\delta\vec{\varphi}$ fluctuations are shifted with respect to the above solutions by an extra $\pm \mEff$ due to the rotating $\vec{e}(t)$. 
This gives rise to a total of 4 positive dispersions given by $M\pm\omega_\text{Bog}$ and $\omega_\text{rad}\pm M$, as well as the corresponding 4 negative branches.

The Bogoliubov excitation is clearly visible in our data presented in the main text, especially in the plots for $N=1$, $N=2$ and $N=3$. In fact, we have checked that all excitations predicted in this calculation can be found in our results for $F$ and $\rho$. However, a detailed discussion of these excitations is beyond the scope of this work and will be the topic of a forthcoming work~\cite{BogusPinPRDinprep}.


\subsection{Out-of-plane excitations: Large-$N$ peak}

Projecting the equation onto the vectors $\ket{b}$ with $b\geq3$ leads instead to
\begin{align}
	&\,2\omega_0 \left[ \sin(\omega_0t)\,\dot\theta_{1b} - \cos(\omega_0t)\,\dot\theta_{2b} \right] \nonumber\\
	&\,- \left[ \cos(\omega_0t)(\partial_t^2-\nabla^2)\theta_{1b} + \sin(\omega_0t)(\partial_t^2-\nabla^2)\theta_{2b} \right] =0.
\end{align}
Fourier transforming with respect to time and momentum gives
\begin{align}
	\left[ \theta_b^A(\tilde\omega+\omega_0) + \theta_b^B(\tilde\omega-\omega_0) \right] \left( \tilde\omega^2 - \mEff^2 - p^2 \right) = 0,
\end{align}
where we defined $\theta_b^A\equiv(\theta_{1b}-i\theta_{2b})/\sqrt{2}$ and $\theta_b^B\equiv(\theta_{1b}+i\theta_{2b})/\sqrt{2}$. From this we can extract
\begin{align}
	\tilde\omega = \pm \sqrt{p^2+\mEff^2}.
\end{align}
The solution is then given by
\begin{align}
	\theta_b^A(\omega) =&\, A_b^+\,\delta(\omega-\omega_0-\sqrt{p^2+\mEff^2}) \nonumber\\
	&\,+ A_b^-\,\delta(\omega-\omega_0+\sqrt{p^2+\mEff^2}), \\
	\theta_b^B(\omega) =&\, B_b^+\,\delta(\omega+\omega_0-\sqrt{p^2+\mEff^2}) \nonumber\\
	&\,+ B_b^-\,\delta(\omega+\omega_0+\sqrt{p^2+\mEff^2}),
\end{align}
where $A^\pm_b$ and $B^\pm_b$ are arbitrary constants.
Taking the rotations of $\vec{e}(t)$ as before into account, we obtain the dispersion for the $\delta\vec{\varphi}$ fluctuations corresponding to the large-$N$ modes of the numerics
\begin{align}
	\omega_\Lor = \sqrt{p^2+\mEff^2}.
\end{align}
Apart from this, we also obtain excitations with dispersions $2\mEff \pm \omega_\Lor$. 
Note that none of these solutions exists for $N=2$.

\vfill

\end{document}